\documentclass[12pt]{article}

\makeatletter

\def\paragraph{\@startsection{paragraph}{4}{\z@}{+2.00ex plus
 +1ex minus +.2ex}{1.5ex plus .2ex}{\it\normalsize}}

\def\section{\@startsection {section}{1}{\z@}{+3.0ex plus +1ex minus
  +.2ex}{2.3ex plus .2ex}{\normalsize\bf}}
\def\subsection{\@startsection{subsection}{2}{\z@}{+2.5ex plus +1ex
minus +.2ex}{1.5ex plus .2ex}{\normalsize\bf}}
\def\subsubsection{\@startsection{subsubsection}{3}{\z@}{+3.25ex plus
 +1ex minus +.2ex}{1.5ex plus .2ex}{\normalsize\bf}}

\expandafter\ifx\csname mathrm\endcsname\relax\def\mathrm#1{{\rm #1}}\fi

\makeatletter

\newcount\@tempcntc
\def\@citex[#1]#2{\if@filesw\immediate\write\@auxout{\string\citation{#2}}\fi
  \@tempcnta\z@\@tempcntb\m@ne\def\@citea{}\@cite{\@for\@citeb:=#2\do
    {\@ifundefined
       {b@\@citeb}{\@citeo\@tempcntb\m@ne\@citea
        \def\@citea{,\penalty\@m\ }{\bf ?}\@warning
       {Citation `\@citeb' on page \thepage \space undefined}}%
    {\setbox\z@\hbox{\global\@tempcntc0\csname
b@\@citeb\endcsname\relax}%
     \ifnum\@tempcntc=\z@ \@citeo\@tempcntb\m@ne
       \@citea\def\@citea{,\penalty\@m}
       \hbox{\csname b@\@citeb\endcsname}%
     \else
      \advance\@tempcntb\@ne
      \ifnum\@tempcntb=\@tempcntc
      \else\advance\@tempcntb\m@ne\@citeo
      \@tempcnta\@tempcntc\@tempcntb\@tempcntc\fi\fi}}\@citeo}{#1}}

\def\@citeo{\ifnum\@tempcnta>\@tempcntb\else\@citea
  \def\@citea{,\penalty\@m}%
  \ifnum\@tempcnta=\@tempcntb\the\@tempcnta\else
   {\advance\@tempcnta\@ne\ifnum\@tempcnta=\@tempcntb \else
\def\@citea{--}\fi
    \advance\@tempcnta\m@ne\the\@tempcnta\@citea\the\@tempcntb}\fi\fi}

\def\asymp#1%
{\mathrel{\raisebox{-.4em}{$\widetilde{\scriptstyle #1}$}}}

\def\Nequal#1%
{\mathrel{\raisebox{-.5em}{$\stackrel{=}{\scriptstyle\rm#1}$}}}

\def\beq{\begin{equation}}
\def\eeq{\end{equation}}
\def\beqar{\begin{eqnarray}}
\def\eeqar{\end{eqnarray}}
\def\barr#1{\begin{array}{#1}}
\def\earr{\end{array}}
\def\bfi{\begin{figure}}
\def\efi{\end{figure}}
\def\btab{\begin{table}}
\def\etab{\end{table}}
\def\bce{\begin{center}}
\def\ece{\end{center}}
\def\nn{\nonumber}
\def\disp{\displaystyle}
\def\text{\textstyle}

\def\al{\alpha}

\def\ga{\gamma}
\def\de{\delta}

\def\eps{\epsilon}

\def\la{\lambda}

\def\si{\sigma}

\def\ieps{\ri\epsilon}

\def\refeq#1{\mbox{(\ref{#1})}}

\def\reffi#1{\mbox{Fig.~\ref{#1}}}

\def\refta#1{\mbox{Table~\ref{#1}}}

\def\refse#1{\mbox{Section~\ref{#1}}}

\def\citere#1{\mbox{Ref.~\cite{#1}}}
\def\citeres#1{\mbox{Refs.~\cite{#1}}}

\def\solid{\raise.9mm\hbox{\protect\rule{1.1cm}{.2mm}}}
\def\dash{\raise.9mm\hbox{\protect\rule{2mm}{.2mm}}\hspace*{1mm}}

\newcommand{\TeV}{\unskip\,\mathrm{TeV}}
\newcommand{\GeV}{\unskip\,\mathrm{GeV}}

\newcommand{\pba}{\unskip\,\mathrm{pb}}

\newcommand{\ri}{{\mathrm{i}}}
\newcommand{\rd}{{\mathrm{d}}}

\newcommand{\Oa}{\mathswitch{{\cal{O}}(\alpha)}}

\newcommand{\M}{{\cal{M}}}

\def\mathswitchr#1{\relax\ifmmode{\mathrm{#1}}\else$\mathrm{#1}$\fi}
\newcommand{\Pf}{\mathswitch  f}
\newcommand{\Pfbar}{\mathswitch{\bar f}}

\newcommand{\PW}{\mathswitchr W}
\newcommand{\Pw}{\mathswitchr w}
\newcommand{\PZ}{\mathswitchr Z}

\newcommand{\Pe}{\mathswitchr e}

\newcommand{\Pd}{\mathswitchr d}
\newcommand{\Pu}{\mathswitchr u}

\newcommand{\Pt}{\mathswitchr t}
\newcommand{\Ptbar}{\mathswitchr{\bar t}}
\newcommand{\Pep}{\mathswitchr {e^+}}
\newcommand{\Pem}{\mathswitchr {e^-}}

\def\mathswitch#1{\relax\ifmmode#1\else$#1$\fi}
\newcommand{\Mf}{\mathswitch {m_\Pf}}
\newcommand{\Mff}{\mathswitch {m_{\Pf'}}}

\newcommand{\MW}{\mathswitch {M_\PW}}

\newcommand{\MZ}{\mathswitch {M_\PZ}}

\newcommand{\Me}{\mathswitch {m_\Pe}}

\newcommand{\sw}{\mathswitch {s_\Pw}}
\newcommand{\cw}{\mathswitch {c_\Pw}}

\def\Li{\mathop{\mathrm{Li}}\nolimits}
\def\Re{\mathop{\mathrm{Re}}\nolimits}

\newcommand{\css}{cross sections}

\newcommand{\AAtt}{\gamma\gamma\to\Pt\Ptbar}
\newcommand{\AAee}{\gamma\gamma\to\Pem\Pep}
\newcommand{\AAff}{\gamma\gamma\to\Pf\Pfbar}
\newcommand{\AAffA}{\gamma\gamma\to\Pf\Pfbar\gamma}
\newcommand{\Born}{{\mathrm{Born}}}

\newcommand{\born}{{\mathrm{Born}}}

\newcommand{\onel}{{\mbox{\scriptsize 1-loop}}}
\newcommand{\weak}{{\mathrm{weak}}}
\newcommand{\cut}{{\mathrm{cut}}}

\newcommand{\unpol}{\mathrm{unpol}}

\newcommand{\QED}{{\mathrm{QED}}}
\newcommand{\CC}{{\mathrm{CC}}}
\newcommand{\NC}{{\mathrm{NC}}}
\newcommand{\virt}{{\mathrm{virt}}}
\newcommand{\soft}{{\mathrm{soft}}}
\newcommand{\coll}{{\mathrm{coll}}}
\newcommand{\sub}{{\mathrm{sub}}}

\newcommand{\sgn}{{\mathrm{sgn}}}
\newcommand{\Ncf}{N_f^{\mathrm{c}}}

\newcommand{\dsl}[1]{\not \hspace{-0.7mm}#1}
\def\dsl{\mathpalette\make@slash}
\def\make@slash#1#2{\setbox\z@\hbox{$#1#2$}%
  \hbox to 0pt{\hss$#1/$\hss\kern-\wd0}\box0}

\def\draftdate{\relax}
\def\mda{\relax}
\def\mua{\relax}
\def\mla{\relax}
\def\draft{
\def\thtystars{******************************}
\def\sixtystars{\thtystars\thtystars}
\typeout{}
\typeout{\sixtystars**}
\typeout{* Draft mode!
         For final version remove \protect\draft\space in source file *}
\typeout{\sixtystars**}
\typeout{}
\def\draftdate{\today}
\def\mua{\marginpar[\boldmath\hfil$\uparrow$]%
                   {\boldmath$\uparrow$\hfil}%
                    \typeout{marginpar: $\uparrow$}\ignorespaces}
\def\mda{\marginpar[\boldmath\hfil$\downarrow$]%
                   {\boldmath$\downarrow$\hfil}%
                    \typeout{marginpar: $\downarrow$}\ignorespaces}
\def\mla{\marginpar[\boldmath\hfil$\rightarrow$]%
                   {\boldmath$\leftarrow $\hfil}%
                    \typeout{marginpar: $\leftrightarrow$}\ignorespaces}
\overfullrule 5pt
\oddsidemargin -15mm
\marginparwidth 29mm
}

\def\stars{\strut\leaders\hbox{*}\hfill\strut}
\def\starline{\hfil\strut\hfil\hbox to \textwidth {\stars}\hfil}

\begin{document}

\thispagestyle{empty}
\def\thefootnote{\fnsymbol{footnote}}
\setcounter{footnote}{1}
\null
\draftdate\hfill CERN-TH/98-335 \\
\strut\hfill PSI-PR-98-28\\
\strut\hfill hep-ph/9812411
\vfill
\begin{center}
{\Large \bf
\boldmath{Production of Light Fermion--Antifermion Pairs \\
in $\gamma\gamma$ Collisions}
\par} \vskip 2.5em
{\large
{\sc A.~Denner%
}\\[1ex]
{\normalsize \it Paul Scherrer Institut\\
CH-5232 Villigen PSI, Switzerland}\\[2ex]
{\sc S.~Dittmaier%
}\\[1ex]
{\normalsize \it Theory Division, CERN\\
CH-1211 Geneva 23, Switzerland}\\[2ex]
}%
\par \vskip 1em%
\end{center}\par
\vskip 2cm
{\bf Abstract:} \par
The ${\cal O}(\alpha)$ corrections to $\AAff$ in the Standard Model 
are calculated for arbitrary, light fermions $f$. The relevant
analytical results are listed in a form that is appropriate for
practical applications, and numerical results for integrated cross
sections are discussed. The corresponding
QED corrections are generally of the order 
of some per mille for arbitrary energies. The weak corrections to $\AAee$ 
are negligible below the electroweak scale, 
reach the per-cent level at a few hundred GeV and grow to about 
$-10\%$ at $2\TeV$. The weak corrections to $\Pu\bar\Pu$ and 
$\Pd\bar\Pd$ production have a shape similar to the one for $\Pem\Pep$, 
but they are larger by factors $\sim 1.4$ and $\sim 3$, respectively.
\par
\vskip 2cm
\noindent
CERN-TH/98-335 \\
December 1998
\null
\setcounter{page}{0}
\clearpage
\def\thefootnote{\arabic{footnote}}
\setcounter{footnote}{0}

\section{Introduction}
\label{se:intro}

Since the suggestion of a photon linear collider (PLC) in the 80's 
\cite{gi83} as an additional option for future $\Pep\Pem$ linear 
colliders, many studies on the feasibility (see \citere{br98} and 
references therein) and the physics potential \cite{br98,br95} of such 
a machine have been performed. A PLC provides an excellent device 
complementary to $\Pep\Pem$ colliders, as can be seen from the
following examples. Photon--photon collisions allow for a search of
Higgs bosons by $s$-channel production and for high-precision tests of
the properties of W bosons, which are produced 
in pairs 
with an enormously large cross section. Moreover, the production cross
sections of charged particles, in many models for new physics, are even
larger than for comparable $\Pep\Pem$ machines \cite{br95}. 
Last but not least, 
a PLC allows various QCD studies, in particular the investigation of
the structure of the photon itself.

According to the DESY/ECFA study \cite{br98} a total $\gamma\gamma$ 
luminosity of $10^{33}\,{\mathrm{cm}}^{-2}\,{\mathrm{s}}^{-1}$, or even 
1--2 orders of magnitude higher, can be reached by Compton backscattering 
of laser photons off the high-energetic $\Pe^\pm$ beams at a $500\GeV$ 
collider. This production mechanism renders the luminosity spectrum
non-trivial, since both photon beams are not monochromatic, and a
luminosity monitor has to be sensitive to both photon energies. 
For this task the processes $\AAee,\mu^-\mu^+$ have been suggested (see 
\citere{br98} and references therein) as reference reactions. Thus,
the lepton-pair production cross section should be known to very high
precision.

In this paper we calculate the complete ${\cal O}(\alpha)$ corrections to
$\AAff$ in the Standard Model for arbitrary light fermions $f$. We present
analytical results that are sufficient for an evaluation of all relevant
observables such as cross sections and distributions
for all polarization configurations.
The structure of the radiative corrections 
and the leading contributions are discussed in detail. 
Moreover, we provide
numerical results on the integrated cross sections and the
corresponding electroweak corrections for the different fermion
flavours.

The paper is organized as follows: in \refse{se:conv&LO} we introduce
some conventions and list analytical results for the 
lowest-order cross sections. 
In \refse{se:EWRC}
the electroweak radiative corrections are classified into QED and weak
corrections, and the corresponding analytical results are presented.
The numerical results are discussed in \refse{se:numres}, and
\refse{se:Sum} contains a summary. Explicit analytical expressions for
the relevant scalar one-loop integrals are given in the appendix.

\section{Conventions and lowest-order cross section}
\label{se:conv&LO}

We consider the reaction
\begin{equation}
\gamma(k_{1},\lambda_{1}) + \gamma(k_{2},\lambda_{2}) \; \longrightarrow \;
\mbox{\Pf}(p,\sigma) + \bar{\mbox{\Pf}}(\bar{p},\bar{\sigma}).
\end{equation}
The mass $\Mf$ of the fermion $f$ is neglected whenever
possible. Otherwise we follow closely the conventions of \citere{de95}.  
The helicities of the incoming photons and of the outgoing fermions
are denoted by $\lambda_{1,2} = \pm 1$ and $\sigma,\bar{\sigma} = \pm
1/2$, respectively.
In the centre-of-mass system (CMS), the momenta read
\begin{eqnarray}
k_1^\mu     &=& \parbox{5cm}{$E(1,0,0,-1),$}
k_2^\mu      =  E(1,0,0,1),  \nn\\
p^\mu       &=& \parbox{5cm}{$E(1,-\sin\theta,0,-\cos\theta),$} 
\bar{p}^\mu  =  E(1, \sin\theta,0, \cos\theta),
\label{eq:AAffmom}
\end{eqnarray}
where $E$ is the energy of the incident photons, and
$\theta$ denotes the scattering angle.
The Mandelstam variables are given by
\begin{eqnarray}
s &=& (k_1+k_2)^2 \;=\; (p+\bar{p})^2 \phantom{_{2}} \;=\; 4 E^2,  \nn\\
t &=& (k_1-p)^2 \phantom{_{2}}  \;=\; (k_2-\bar{p})^2
  \;=\; -4E^2\sin^2\text\frac{\theta}{2}, \nn\\
u &=& (k_1-\bar{p})^2 \phantom{_{2}} \;=\; (k_2-p)^2
  \;=\; -4E^2\cos^2\text\frac{\theta}{2}.
\label{eq:stu}
\end{eqnarray}
The neglect of $\Mf$ in the kinematics implies that our results are valid
for $s,-t,-u\gg\Mf^2$. 

The scattering amplitude of $\AAff$ obeys Bose symmetry
with respect to the incoming photons
and---neglecting quark mixing---also CP symmetry. 
Consequently, the polarized cross sections
$\rd\sigma^{\la_1,\la_2,\si,\bar{\si}}$ are related by
\begin{eqnarray}
\rd\sigma^{\la_1,\la_2,\si,\bar{\si}}(s,t,u) 
&=& \parbox{4.5cm}{$\rd\sigma^{\la_2,\la_1,\si,\bar{\si}}(s,u,t)$}
(\mbox{Bose}) \nn\\
&=& \parbox{4.5cm}{$\rd\sigma^{-\la_1,-\la_2,-\bar\si,-\si}(s,u,t)$}
(\mbox{CP}) \nn\\
&=& \parbox{4.5cm}{$\rd\sigma^{-\la_2,-\la_1,-\bar\si,-\si}(s,t,u)$}
(\mbox{Bose + CP}).
\label{eq:BoseCP}
\end{eqnarray}
In lowest order, $\gamma\gamma\to\Pf\Pfbar$ is a pure QED process
and is therefore invariant under parity.
Hence, the Born cross sections obey the additional relations
\beq
\rd\sigma_\Born^{\la_1,\la_2,\si,\bar{\si}}(s,t,u) \;=\;
\rd\sigma_\Born^{-\la_1,-\la_2,-\si,-\bar{\si}}(s,t,u) \qquad\quad
(\mbox{P}).
\eeq

\begin{figure}
\begin{center}
\begin{picture}(10,2.6)
\put(-4.0,-14.4){\includegraphics{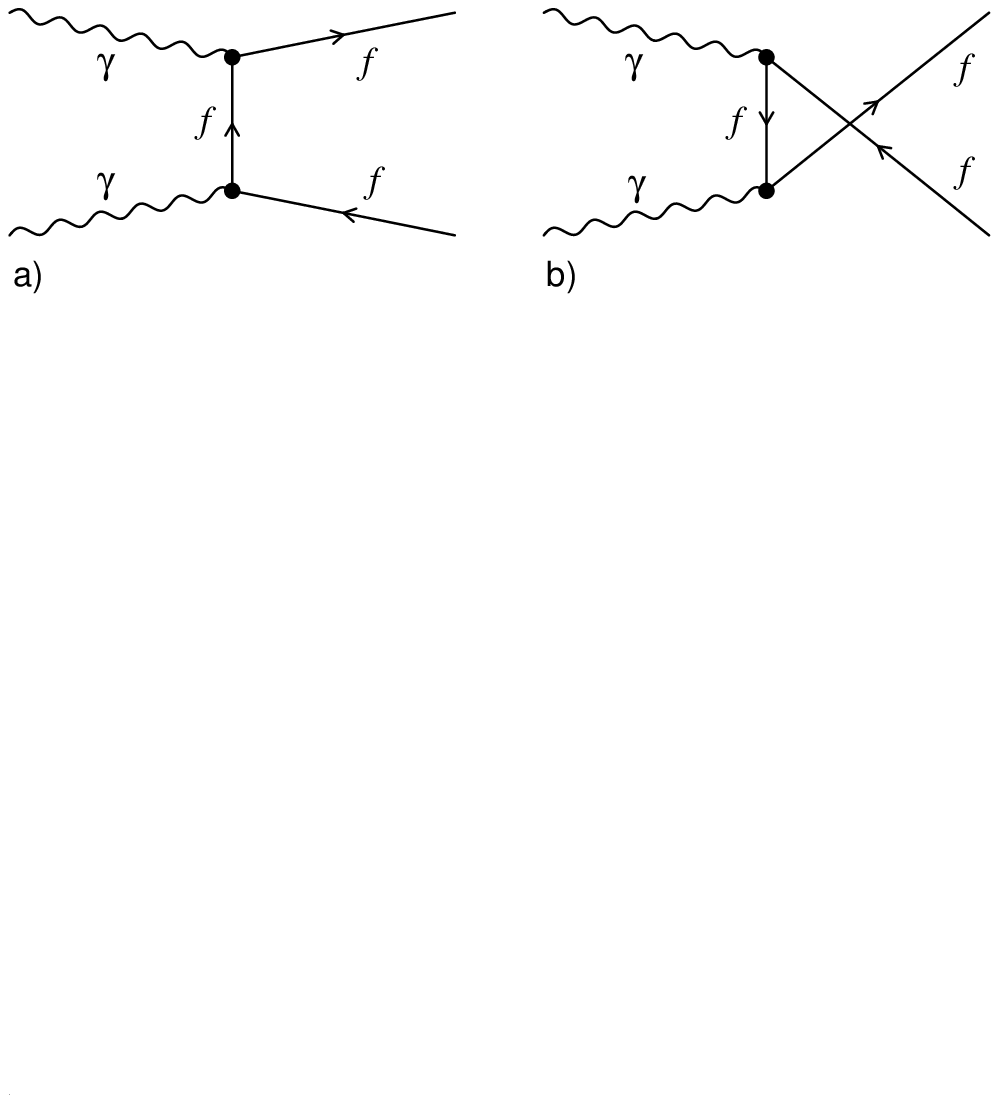}}
\end{picture} 
\end{center}
\caption{Tree diagrams for $\AAff$}
\label{fig:lodiags}
\end{figure}

The two lowest-order Feynman diagrams%
\footnote{All Feynman diagrams in this work have been drawn with the
help of {\sl FeynArts} \cite{fa}.}
are shown in \reffi{fig:lodiags}. 
The differential Born cross section reads
\beq
\frac{\rd\sigma_\born}{\rd\Omega}(P_1,P_2) =
\frac{\Ncf}{64\pi^2 s} \sum_{\lambda_1,\lambda_2,\sigma,\bar\sigma}
\frac{1}{4}(1+\lambda_1 P_1)(1+\lambda_2 P_2)
|\M_\born^{\lambda_1,\lambda_2,\sigma,\bar\sigma}(s,t,u)|^2,
\eeq
where $P_{1,2}$ are the degrees of beam polarization, and the sum on the
r.h.s.\ includes the desired polarizations of the outgoing particles.
The colour factor for the fermion $f$ is denoted by $\Ncf$, i.e.\
$N^{\mathrm{c}}_{\mathrm{lepton}}=1$ and $N^{\mathrm{c}}_{\mathrm{quark}}=3$.
The squares of the helicity amplitudes
$\M_\born^{\lambda_1,\lambda_2,\sigma,\bar\sigma}$ are given by
\beqar
|\M_\born^{\lambda_1,\lambda_2,\sigma,\bar\sigma}(s,t,u)|^2 &=& \left\{
\barr{cl}
\disp 4Q_\Pf^4 e^4\frac{u}{t} \quad
& \mbox{for}\quad \lambda_1=-\lambda_2=\pm1, \;
\sigma=-\bar\sigma=\pm\frac{1}{2}, \\[.8em]
\disp 4Q_\Pf^4 e^4\frac{t}{u} \quad
& \mbox{for}\quad \lambda_1=-\lambda_2=\mp1, \;
\sigma=-\bar\sigma=\pm\frac{1}{2}, \\[.8em]
0 & \mbox{otherwise.}
\earr\right.
\label{eq:mborn}
\eeqar
The $t$- and $u$-channel poles in the squared amplitudes lead to 
kinematical singularities in the very 
forward and backward directions, where
we are not interested in the \css, since the fermions escape into the beam
pipe. For leptons, these singularities are of course regulated by a finite 
lepton mass. For light-quark production in the forward and backward
directions purely perturbative calculations are not reliable, since the
splitting of a photon into a nearly collinear quark--antiquark pair
involves QCD effects at very low scales.
We avoid the forward and backward regions by imposing the angular cut
\beq
\theta_\cut<\theta<180^\circ-\theta_\cut.
\label{eq:aaffcut}
\eeq
For later convenience, we introduce the step function
\beq
g_\cut(\theta) = \Theta(\theta-\theta_\cut)
\Theta(180^\circ-\theta_\cut-\theta),
\label{eq:step}
\eeq
where $\Theta(x)$ is the usual Heaviside distribution.
Integrating over a symmetric angular range \refeq{eq:aaffcut}, the 
contributions of all non-vanishing Born \css\ are equal, and the 
integrated, unpolarized cross section reads
\beqar
\si_\born^\unpol = \Ncf Q_\Pf^4\alpha^2 \frac{4\pi}{s}
\left[ \ln\left(\frac{1+\cos\theta_\cut}{1-\cos\theta_\cut}\right)
-\cos\theta_\cut \right],
\label{eq:intborn}
\eeqar
where $\alpha=e^2/(4\pi)$ is the fine-structure constant.

\section{Electroweak radiative corrections}
\label{se:EWRC}

\subsection{Classification of \boldmath{${\cal O}(\alpha)$} corrections
and general remarks}

\begin{figure}
\begin{center}
\begin{picture}(16,3.7)
\put(-3.8,-13.7){\includegraphics{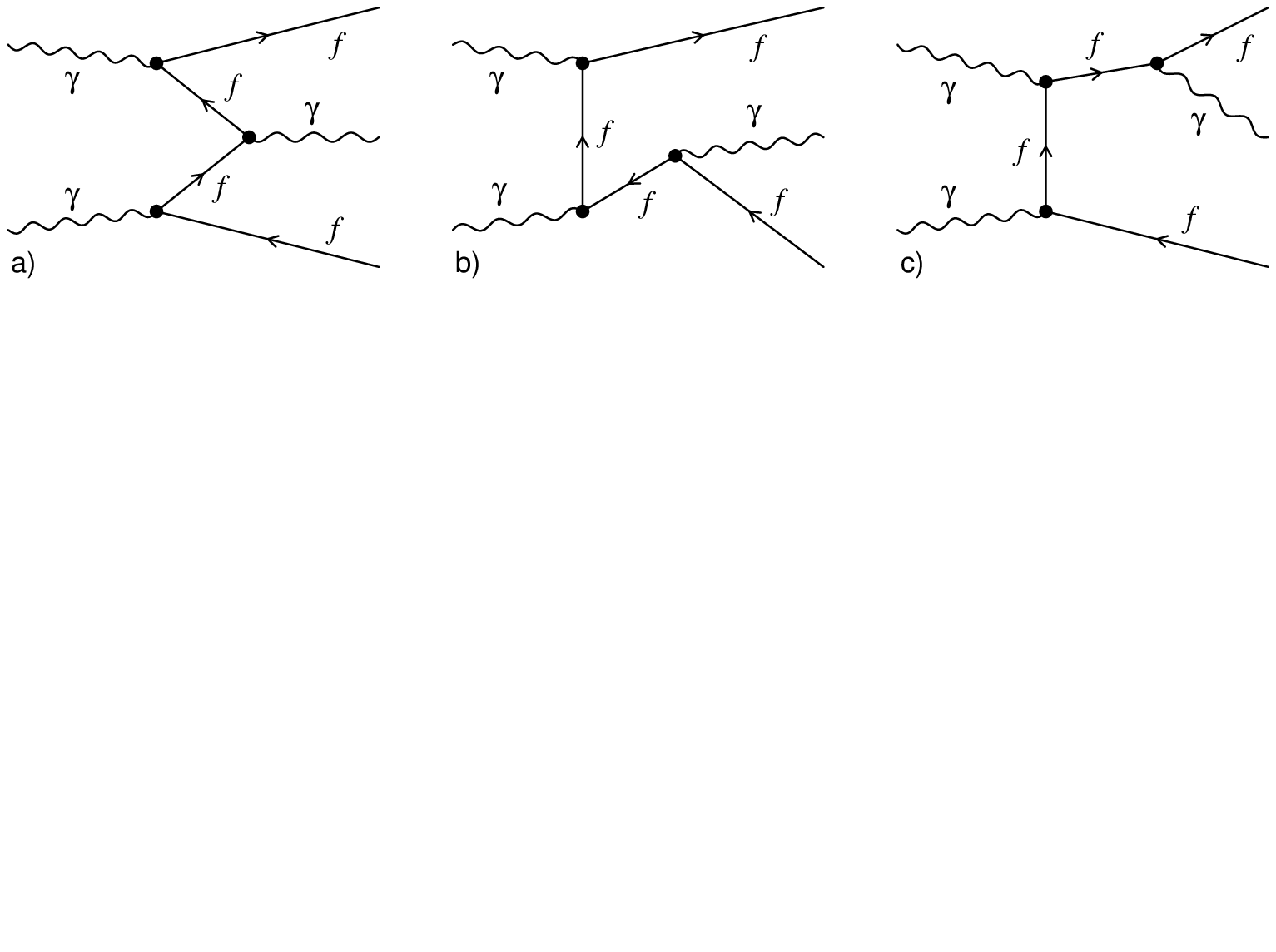}}
\put( 6.0,-0.2){+ crossed graphs}
\end{picture} 
\end{center}
\caption{Diagrams for photon bremsstrahlung in $\AAff$ 
(``crossing'' means interchanging the incoming photon lines.)}
\label{fig:bremdiags}
\end{figure}

\begin{figure}
\begin{center}
\begin{picture}(16,3.0)
\put(-3.8,-17.2){\includegraphics{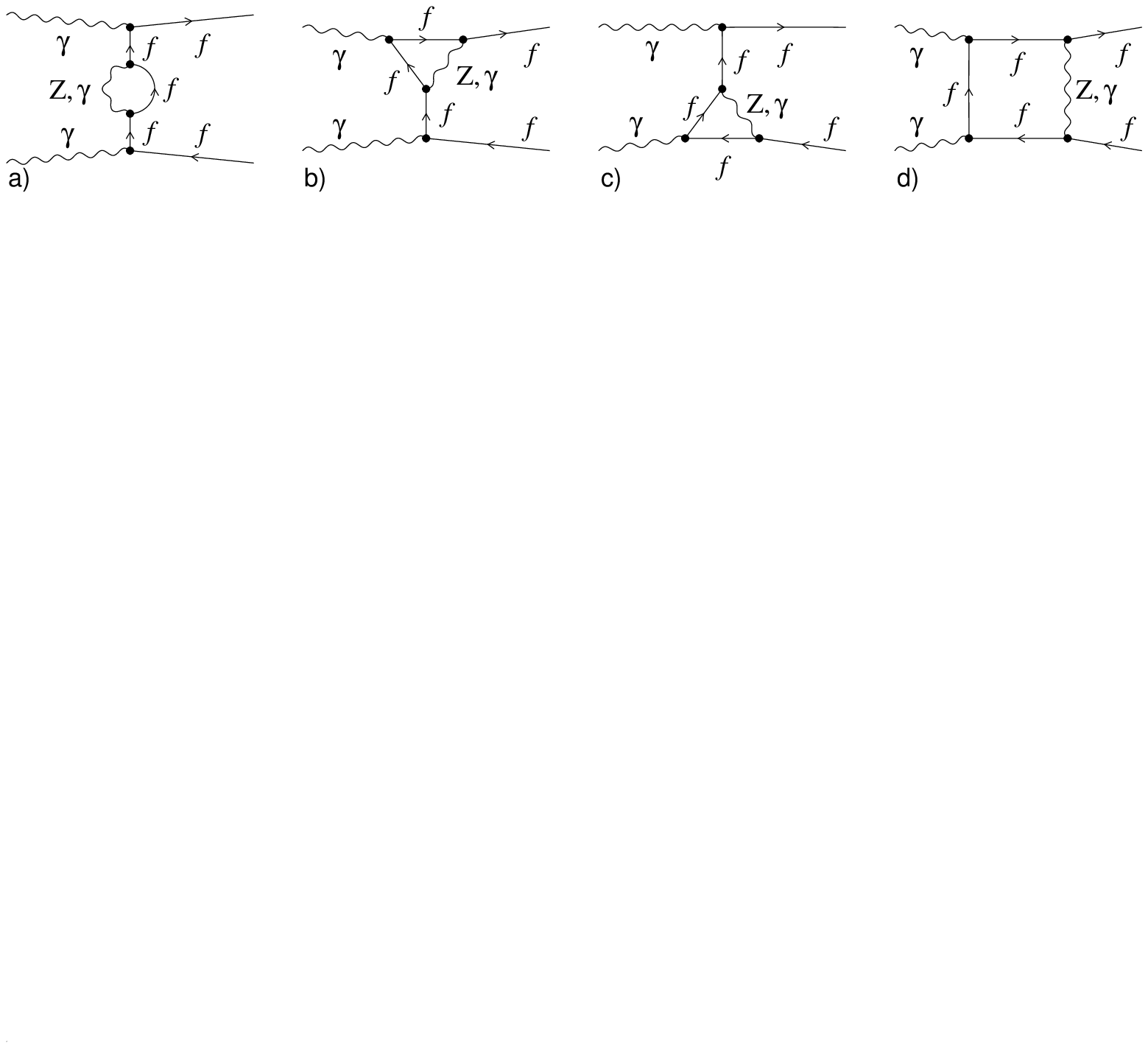}}
\put( 6.0,-0.2){+ crossed graphs}
\end{picture} 
\end{center}
\caption{Diagrams for $\AAff$ with virtual \PZ-boson or photon 
exchange} 
\label{fig:zadiags}
\end{figure}

\begin{figure}
\begin{center}
\begin{picture}(16,8.0)
\put(-3.8,-12.2){\includegraphics{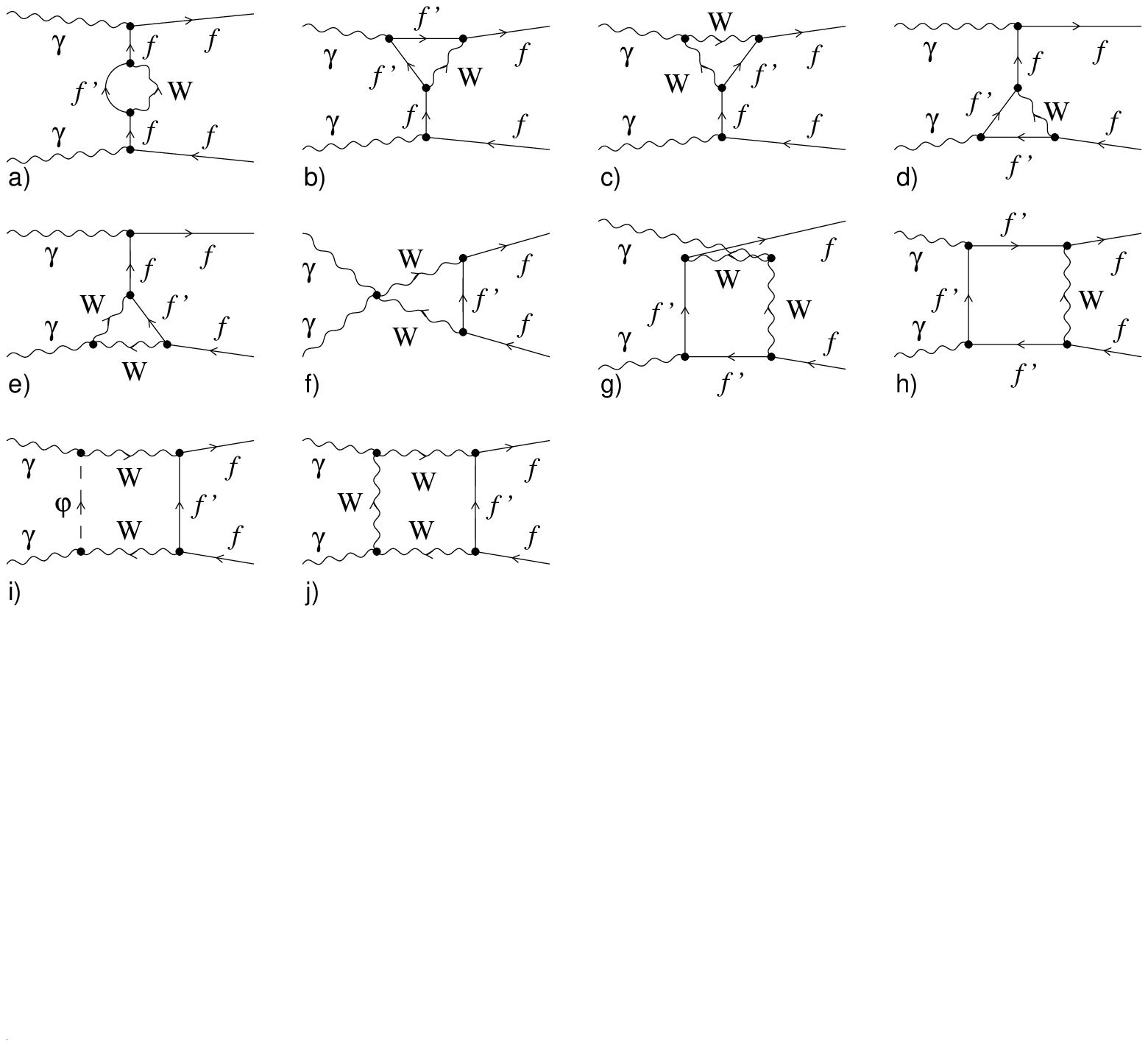}}
\put( 9.0,  1.0){+ crossed graphs [except for graph f)]}
\end{picture}
\end{center}
\caption{Diagrams for $\AAff$ with virtual \PW-boson exchange}
\label{fig:wdiags}
\end{figure}

Since $\gamma\gamma\to\Pf\Pfbar$ is a pure QED process in lowest order,
the SM electroweak corrections of ${\cal O}(\alpha)$ consist of two
separately gauge-invariant types: pure QED corrections and genuinely
weak corrections. The QED corrections include real photon emission (see
\reffi{fig:bremdiags}), virtual photon exchange (see \reffi{fig:zadiags}), 
and the corresponding 
counterterms. The weak corrections comprise all one-loop
diagrams (and contributions to counterterms) that involve the massive
weak gauge bosons W and Z. For vanishing fermion mass $\Mf$, there are
no contributions 
involving Higgs-boson exchange or closed fermion loops.%
\footnote{There are actually Feynman diagrams involving fermion-loop
  contributions to the $AAZ^*$, $AA\chi^*$, and $AAH^*$ vertices. 
  However, these contributions are proportional to the mass of the
  produced fermion $f$ \cite{de95} and are thus neglected.}
As a consequence the $\Oa$ corrections do not depend on the
Higgs-boson and top-quark masses and on the running of $\al$.  
The weak corrections can be further classified into two
subsets%
\footnote{In the $R_\xi$ gauges these subsets are gauge-independent.}. 
The first of these subsets includes all
diagrams that contain internal Z-boson lines (see
\reffi{fig:zadiags}), and the corresponding corrections are called
{\it neutral-current} (NC) corrections in the following. The second
subset includes all diagrams with W-boson exchange (see
\reffi{fig:wdiags}), leading to {\it charged-current} (CC)
corrections. Note that only the CC corrections involve non-abelian
couplings among the gauge bosons.

The perturbative QCD corrections can be obtained 
from the QED corrections 
by substituting the electromagnetic coupling factor $Q_f^2\alpha$ by 
the strong coupling factor $4\alpha_{\mathrm{s}}/3$.
The definition of a proper two-jet cross section is, however, problematic, 
because jets of radiated gluons cannot be distinguished from those
of quarks. Consequently, a two-jet cross section 
includes the case where a gluon
and one of the quarks cause the two jets, whereas one of the quarks
disappears in the beam pipe. This contribution is divergent,
owing to the $t$- or $u$-channel poles that are 
not cut out by this definition of two-jet events.
Therefore, a consistent definition of two-jet events in $\gamma\gamma$
collisions necessarily involves non-perturbative effects,
which will not be discussed in this paper.

The renormalization of the ${\cal O}(\alpha)$ corrections turns out to
be extremely simple for $\AAff$. For $\Mf=0$, the mass renormalization
drops out, and only the wave-function renormalization of the external 
fields and the charge renormalization are relevant. Note that the
photonic wave-function renormalization constant exactly cancels against
a corresponding part in the charge renormalization (see for instance
\citere{de93}) 
so that no effects from the photonic vacuum polarization remain.
Consequently, there is no running in the
electromagnetic coupling $\alpha$ for $\AAff$ in this order. The
remaining part of the charge renormalization is the contribution of
the photon--Z-boson-mixing self-energy at zero-momentum transfer, which in
the usual 't~Hooft--Feynman gauge consists of a W-boson loop only and
is thus part of the CC corrections.

Virtual one-loop
corrections are included into predictions by replacing the squared Born
amplitude $|\M_\born|^2$ by $|\M_\born|^2+2\Re\{\M_\onel\M_\born^*\}$,
where $\M_\onel$ is the contribution of the one-loop diagrams to the 
scattering amplitude. Thus, the one-loop correction to a lowest-order
cross section is zero whenever the lowest-order vanishes, and we can
factorize the one-loop correction $\rd\sigma_\onel$ to the differential
cross section into the lowest-order cross section $\rd\sigma_\born$ 
and the relative correction $\delta_\onel$ for each polarization
configuration:
\beq
\rd\sigma_\onel = \delta_\onel \rd\sigma_\born, \qquad
\delta_\onel = 
\delta_\NC + \delta_\CC + \delta^\virt_\QED.
\label{eq:onel}
\eeq
According to the above decomposition, $\delta_\onel$ is split into NC, 
CC, and QED contributions. Since the Born amplitudes 
are non-vanishing only for $\lambda_1=-\lambda_2$ and $\sigma=-\bar\sigma$
[see \refeq{eq:mborn}],
we introduce $\rho=\sgn(\lambda_1)=-\sgn(\lambda_2)$ and
$\kappa=\sgn(\sigma)=-\sgn(\bar\sigma)$,
and indicate the polarization 
configurations for the relative corrections $\delta_{\dots}$ in 
\refeq{eq:onel} by $\delta^{\rho,\kappa}_{\dots}$.

The calculation of the one-loop diagrams has been
performed by applying the standard techniques summarized in 
\citere{de93}. More precisely, tensor one-loop integrals are
algebraically reduced to scalar integrals, as described in \citere{pa79},
and scalar integrals are computed using the methods and results of
\citere{th79}. Technically, the algebraic evaluation of the Feynman
amplitudes, which have been generated with {\sl FeynArts} \cite{fa}, has
been carried out in the same way as described in \citere{de95} for
$\AAtt$. In
particular, the algebraic manipulations have been performed again twice,
once using {\sl FeynCalc} \cite{fc} and once using our 
own {\sl Mathematica}
\cite{math} routines. For $\AAee$ the virtual corrections are related to
the ones of $\Pem\gamma\to\Pem\gamma$ \cite{di94} by crossing symmetry,
which has been used as additional check for this channel.

The evaluation of the real-photonic bremsstrahlung will be described
in detail below. 

\subsection{Weak corrections}
\label{se:weak}

The NC corrections for the different polarization channels are related by 
Bose and parity transformations as follows:
\beq
\delta_{\NC}^{\rho,\kappa} =
\delta_{\NC}^{-\rho,\kappa}\Big|_{t\leftrightarrow u}
= \delta_{\NC}^{-\rho,-\kappa}
\left(g_{\Pf\Pf\PZ}^\kappa/g_{\Pf\Pf\PZ}^{-\kappa}\right)^2,
\label{eq:NCsym}
\eeq
where $g_{\Pf\Pf\PZ}^\kappa$ is the generic \PZ\Pf\Pfbar\ coupling,
\beq
g_{\Pf\Pf\PZ}^\kappa=-\frac{\sw}{\cw}Q_\Pf 
+ \frac{I^3_\Pf}{\sw\cw}\delta_{\kappa-},
\eeq
with $I^3_\Pf=\pm\frac{1}{2}$ denoting the weak isospin of the
left-handed component of the fermion $f$. 
The cosine $\cw$ of the weak
mixing angle is fixed by the ratio of the masses $\MW$ and $\MZ$ of the
weak gauge bosons, i.e.\ $\cw^2=1-\sw^2=\MW^2/\MZ^2$.
According to \refeq{eq:NCsym}, all NC correction factors can be deduced from 
\beqar
\delta_{\NC}^{+,-} &=& 
\frac{\alpha}{\pi} \left(g_{\Pf\Pf\PZ}^-\right)^2 
\left\{
\left(1-\frac{\MZ^2}{u}\right) 
\left(\frac{3}{2}+\frac{u}{t}+\frac{\MZ^2}{2u}-\frac{\MZ^2}{t}\right) 
\ln\left(1-\frac{u}{\MZ^2}\right) 
-\frac{u}{t}\ln\left(\frac{s}{\MZ^2}\right)
\right. \nn\\[.2em]
&& 
\left. {}
+\frac{(s+\MZ^2)^2}{t^2} \left[ 
\ln\left(\frac{s}{\MZ^2}\right)\ln\left(\frac{\MZ^2+s}{\MZ^2-u}\right)
+\Li_2\left(-\frac{s}{\MZ^2}\right) 
-\Li_2\left(\frac{u}{\MZ^2}\right) 
\right] \right. \nn\\[.2em]
&& 
\left. {}
-\frac{(t-\MZ^2)^2}{t^2} \left[
\ln\left(\frac{s}{\MZ^2}\right)
\ln\left(1-\frac{t}{\MZ^2}\right) 
+\Li_2\left(\frac{t}{\MZ^2}\right) \right] 
+ \frac{\MZ^2}{t} - \frac{\MZ^2}{2u} - \frac{5}{4} 
\right\}.
\label{eq:ncrc}
\eeqar
Note that the contributions to the one-loop correction $\de_\onel$ 
\refeq{eq:onel} are real quantities; the imaginary parts of the one-loop
integrals do not contribute.

The CC corrections vanish for right-handed fermions,
and the corrections for left-handed fermions are related by Bose
symmetry,
\beq
\delta_\CC^{\rho,+} = 0, \qquad
\delta_{\CC}^{+,-} = \delta_{\CC}^{-,-}\Big|_{t\leftrightarrow u}.
\eeq
For $\delta_{\CC}^{-,-}$ we explicitly obtain
\beqar\label{eq:dCC}
\delta_{\CC}^{-,-} &=& \frac{\alpha}{4\pi\sw^2}  
\Re\Biggr\{ 
\frac{1}{2}
-\frac{3tu+\MW^2u-2\MW^2t}{ut}B_\Pw(t)
\nn\\[.2em] && {}
+\frac{(Q_{f'}-Q_f)}{Q_f}\,\frac{2(\MW^2-u)^2}{u^2}
\left[t\bar C_{\Pw\Pw}(t)+u\bar C_{\Pw\Pw}(u)\right]
\nn\\[.2em] && {}
+\frac{(Q_f-Q_{f'})^2}{Q_f^2}\,\Biggl(
\frac{2t}{u}\left[B_\Pw(t)-B_{\Pw\Pw}(s)\right]
+\frac{st(t+2\MW^2)(t-u)}{u^2}D_{\Pw\Pw\Pw}(s,t)
\nn\\[.2em]
&& \hspace{3em} {}
+\frac{t(u-t-2\MW^2)}{u^2}
\left[-s\MW^2 D_{\Pw\Pw\Pw}(s,t)
+sC_{\Pw\Pw\Pw}(s)+2t\bar C_{\Pw\Pw}(t)+sC_{\Pw\Pw}(s)\right]
\nn\\[.2em]
&& \hspace{3em} {}
+\frac{(\MW^2-u)^2}{u^2}
\left[(s\MW^2-st-2t^2)D_{\Pw\Pw\Pw}(s,t)\right.
\nn\\[.2em]
&& \hspace{9em} \left. {}
+(s\MW^2-su-2u^2)D_{\Pw\Pw\Pw}(s,u)-2sC_{\Pw\Pw\Pw}(s)\right]
\Biggr)
\nn\\[.2em] && {}
+\frac{Q_{f'}(Q_{f'}-Q_f)}{Q_f^2}\,\frac{2(\MW^2-u)^2}{u^2}
\left[(tu+s\MW^2)D_{\Pw\Pw}(u,t)-t\bar C_\Pw(t)-u\bar C_\Pw(u)\right]
\nn\\[.2em] && {}
+\frac{Q_{f'}^2}{Q_f^2}\,\Biggl(
\frac{2t}{u} \left[B(s)-B_\Pw(t)\right]
-\frac{st(2\MW^2-t-2u)}{u^2}C_\Pw(s)
\nn\\[.2em]
&& \hspace{3em} {}
+\frac{(s+\MW^2)^2}{u^2}
\left[s(\MW^2-t)D_\Pw(s,t)+sC(s)+2t\bar C_\Pw(t)\right]
\nn\\[.2em]
&& \hspace{3em} {}
+\frac{(\MW^2-u)^2}{u^2}
\left[s(\MW^2-u)D_\Pw(s,u)+sC(s)+2u\bar C_\Pw(u)\right]
\Biggr)
\; \Biggl\}, 
\label{eq:ccrc}
\eeqar
where $Q_{f'}=Q_f-2I^3_f$ denotes the charge of the weak isospin partner
$f'$ of the fermion $f$.
The functions $B_{\dots}$, $C_{\dots}$, and $D_{\dots}$ 
are scalar one-loop integrals, the explicit expressions of which are
collected in the appendix. Although some of the scalar integrals contain
(logarithmic) mass singularities, which are regularized by infinitesimal
masses $\Mf$ and $\Mff$, all mass singularities drop out in the final
results for $\delta_\CC$.

Note that the relative weak corrections vanish directly on the $t$- and
$u$-channel poles of the lowest-order cross sections, i.e.\ the weak
corrections are suppressed where the differential cross section is maximal. 
This is a consequence of the usual charge renormalization, which defines 
the charge $e=\sqrt{4\pi\alpha}$ as the $\gamma f\bar f$ coupling for
all particles on shell.
Since this kinematic situation holds for forward and backward
scattering, and since the weak box diagrams do not develop $t$- and
$u$-channel poles, all weak corrections are absorbed by the corresponding
renormalization terms
in this special situation.

\subsection{QED corrections}
\label{se:QED}

\subsubsection{Virtual corrections}
\label{se:QEDvirt}

Bose and P symmetry relate the virtual QED corrections by
\beq
\delta_{\QED}^{\mathrm{virt},\rho,\kappa} =
\left.\delta_{\QED}^{\mathrm{virt},-\rho,\kappa}\right|_{t\leftrightarrow u}
= \delta_{\QED}^{\mathrm{virt},-\rho,-\kappa},
\eeq
so that it is sufficient to give one particular polarization
configuration.
Introducing an infinitesimal photon mass $\lambda\ll\Mf$ as IR regulator, and
keeping $\Mf$ in the mass-singular terms, we obtain 
\beqar
\delta_{\QED}^{\mathrm{virt},+,-} &=& Q_\Pf^2\frac{\alpha}{\pi} 
\left\{
\ln\left(\frac{-t}{\lambda^2}\right)
\left[ 1+\ln\left(\frac{\Mf^2}{s}\right) \right]
-\frac{1}{2}\ln\left(\frac{\Mf^2}{-u}\right)
+\frac{1}{2}\ln^2\left(\frac{\Mf^2}{-t}\right) \right.
\nn\\[.2em] && 
\left. {}
+\ln\left(-\frac{s}{t}\right)
+\frac{s}{t}\ln\left(-\frac{s}{u}\right)
+\frac{s^2}{2t^2}\ln^2\left(-\frac{s}{u}\right)
-\frac{3}{2}
+\frac{2\pi^2}{3} \right\}. 
\label{eq:qedrc}
\eeqar
The IR divergence drops out after adding the real-photonic
bremsstrahlung corrections, 
and the mass singularities cancel completely against mass-singular real
corrections caused by collinear photon emission, since only final-state
particles radiate off photons.

\subsubsection{Real corrections}
\label{se:QEDreal}

Real photon emission in $\AAff$ leads to the kinematically different
process
\beq
\gamma(k_1,\lambda_1) + \gamma(k_2,\lambda_2) \; \longrightarrow \;
f(p,\sigma) + \bar f(\bar p,\bar\sigma) + \gamma(k',\lambda'),
\eeq
with $k'$ and $\lambda'$ denoting the momentum and helicity of the
radiated photon, respectively. While the incoming momenta $k_{1,2}$ are 
the same as for $\AAff$, as specified in \refeq{eq:AAffmom}, 
in the CMS, the outgoing momenta read 
\beqar
p^\mu &=& E_f(1,-\cos\phi_f\sin\theta_f,-\sin\phi_f\sin\theta_f,
-\cos\theta_f),
\nn\\
\bar p^\mu &=& E_{\bar f}(1,\cos\phi_{\bar f}\sin\theta_{\bar f},
\sin\phi_{\bar f}\sin\theta_{\bar f},\cos\theta_{\bar f}),
\nn\\
k^{\prime \mu} &=& E'(1,\cos\phi'\sin\theta',\sin\phi'\sin\theta',
\cos\theta').
\eeqar
The lowest-order cross section for $\AAffA$, which yields an 
${\cal O}(\alpha)$ correction to $\AAff$, is given by
\beq
\sigma_\gamma(P_1,P_2) =
\frac{\Ncf}{2s} \int \rd\Gamma \,
\sum_{\lambda_1,\lambda_2,\sigma,\bar\sigma,\lambda'}
\frac{1}{4}(1+\lambda_1 P_1)(1+\lambda_2 P_2) \,
|\M_\gamma^{\lambda_1,\lambda_2,\sigma,\bar\sigma,\lambda'}|^2,
\label{eq:hbcs}
\eeq
where the phase-space integral is defined by
\beq
\int \rd\Gamma =
\int\frac{\rd^3 {\bf p}}{(2\pi)^3 2E_f}
\int\frac{\rd^3 {\bf \bar p}}{(2\pi)^3 2E_{\bar f}}
\int\frac{\rd^3 {\bf k}'}{(2\pi)^3 2E'} \,
(2\pi)^4 \delta(k_1+k_2-p-\bar p-k).
\eeq

We have calculated the helicity amplitudes $\M_\gamma$ in two different
ways. One calculation is performed by applying the 
Weyl--van-der-Waerden spinor technique (see \citere{di98} and references
therein); the second calculation makes use of an explicit representation
of spinors, polarization vectors, and Dirac matrices.
For $\Mf=0$ the helicity structure forces many helicity amplitudes to
vanish. In particular, $\M_\gamma$ is zero if $\sigma=\bar\sigma$
or $\lambda_1=\lambda_2=-\lambda'$. Moreover, Bose, CP, P, and
crossing symmetry for in- and outgoing photons lead to relations among
the helicity amplitudes. In order to be independent of phase
conventions, we formulate these relations for $|\M_\gamma|^2$,
\beqar
|\M_\gamma^{\lambda_1,\lambda_2,\sigma,\bar\sigma,\lambda'}|^2 
&=& \parbox{7.5cm}{$
|\M_\gamma^{\lambda_2,\lambda_1,\sigma,\bar\sigma,\lambda'}|^2 
\Big|_{k_1\leftrightarrow k_2} $} \mbox{(Bose)}
\nn\\
&=& \parbox{7.5cm}{$
|\M_\gamma^{-\lambda_1,-\lambda_2,-\bar\sigma,-\sigma,-\lambda'}|^2 
\Big|_{p\leftrightarrow \bar p} $} \mbox{(CP)}
\nn\\
&=& \parbox{7.5cm}{$
|\M_\gamma^{-\lambda_1,-\lambda_2,-\sigma,-\bar\sigma,-\lambda'}|^2 
$} \mbox{(P)}
\nn\\
&=& \parbox{7.5cm}{$
|\M_\gamma^{-\lambda',\lambda_2,\sigma,\bar\sigma,-\lambda_1}|^2 
\Big|_{k_1\leftrightarrow -k'} $} \mbox{(crossing)}.
\eeqar
Owing to these relations, only one independent non-vanishing
helicity amplitude is left, for which we take
\beq
|\M_\gamma^{{+},{-},{-},{+},{+}}|^2 = 4Q_f^6 e^6 \,
\frac{(p\cdot k_1)^2(p\cdot\bar p)}
{(p\cdot k_2)(p\cdot k')(\bar p\cdot k_2)(\bar p\cdot k')}.
\eeq
>From this particular $|\M_\gamma|^2$ we can read off all different kinds
of singularities that can occur for $\AAffA$. Firstly, there are
collinear poles if $f$ or $\bar f$ are scattered into forward or backward
directions, similar to forward or backward scattering in $\AAff$. In
this case we again apply angular cuts in order to exclude that $f$ or
$\bar f$ escape into the beam pipe, i.e.\ we assume 
\beq
\theta_\cut<\theta_f,\theta_{\bar f}<180^\circ-\theta_\cut.
\label{eq:aaffacut}
\eeq
Secondly, we encounter the usual soft and collinear singularities if
$k'$ becomes soft or collinear to $p$ or $\bar p$. These singularities
are the counterparts to the IR and mass singularities in the virtual
QED corrections given in \refse{se:QEDvirt};
they have to be regularized,
as in the virtual case, by the infinitesimal photon mass $\lambda$ and
the fermion mass $\Mf$.
In the following we describe three different procedures for the
treatment of these singularities.

\paragraph{IR phase-space slicing and effective collinear factors}

In order to apply phase-space slicing to the IR singularity, we
exclude the region $E'<\Delta E$ from phase space so that the
IR singularity is regularized by the cut energy $\Delta E\ll E$. 
In the soft-photon region $\lambda<E'<\Delta E$, the asymptotic form of 
the exact differential cross section 
is known to factorize into the lowest-order cross section
without photon emission and a universal eikonal factor, which depends
on the photon momentum (see e.g.\ \citere{de93}). The integration over
the soft-photon phase space, which is carried out in the CMS, 
yields the simple correction factor $\delta_\soft$ to the
differential Born cross section $\rd\sigma_\born$ for $\AAff$:
\beq
\delta_\soft = -Q_f^2 \frac{\alpha}{\pi} \left\{
2\ln\left(\frac{2\Delta E}{\lambda}\right)
\left[ 1+\ln\left(\frac{\Mf^2}{s}\right) \right]
+\frac{1}{2}\ln^2\left(\frac{\Mf^2}{s}\right)
+\ln\left(\frac{\Mf^2}{s}\right)+\frac{\pi^2}{3} \right\}.
\eeq
The factor $\delta_\soft$ does not depend on the polarizations of the
produced fermions and of
the incoming photons, and its dependence on $\la$
obviously cancels against the one in $\delta_\QED^\virt$ given in 
\refse{se:QEDvirt}.

The remaining phase-space integration in \refeq{eq:hbcs} with $E'>\Delta E$ 
still contains the collinear singularities in the regions in which 
$(p\cdot k')$ or $(\bar p\cdot k')$ is small. In these regions,
however, the asymptotic behaviour of the differential cross section 
(including its dependence on $\Mf$) has
a well-known form (see e.g.\ \citere{be82}). The singular terms are
universal and factorize from $\rd\sigma_\born$. A simple
approach to include the collinear regions consists in a suitable
modification of $|\M_\gamma|^2$, which was calculated for $\Mf=0$. 
More precisely, $|\M_\gamma|^2$ is multiplied by an {\it effective
collinear factor} that is equal to 1 up to terms of ${\cal O}(\Mf^2/s)$
outside the collinear regions, but replaces the poles in $(p\cdot k')$
and $(\bar p\cdot k')$ by the correctly mass-regularized behaviour.
Explicitly, the described substitution reads
\beqar
\sum_{\lambda'=\pm 1}
|\M_\gamma^{\lambda_1,\lambda_2,\sigma,\bar\sigma,\lambda'}|^2 
& \;\to\; &
\sum_{\tau,\bar\tau=\pm 1} 
f_\tau(x_f,E_f,\alpha_f) f_{\bar\tau}(x_{\bar f},E_{\bar f},\alpha_{\bar f})
\nn\\
&& {} \times \sum_{\lambda'=\pm 1}
|\M_\gamma^{\lambda_1,\lambda_2,\tau\sigma,\bar\tau\bar\sigma,\lambda'}|^2.
\eeqar
The functions $f_\pm$ describe collinear photon emission with and
without spin flip of the radiating fermion,
\beqar
f_+(x_f,E_f,\alpha_f) & = &
\left(\frac{4E_f^2\sin^2(\frac{\alpha_f}{2})}
{4E_f^2\sin^2(\frac{\alpha_f}{2})+\Mf^2}\right)^2,
\nn\\
f_-(x_f,E_f,\alpha_f) & = &
\frac{x_f^2}{x_f^2+2x_f+2} \frac{4\Mf^2 E_f^2\sin^2(\frac{\alpha_f}{2})}
{[4E_f^2\sin^2(\frac{\alpha_f}{2})+\Mf^2]^2},
\qquad x_f=\frac{E'}{E_f},
\eeqar
where $\alpha_f=\angle({\bf k}_f,{\bf k}')$ is the angle of the photon
emission from $f$. The functions $f_\pm$ describing photon emission from 
$\bar f$ follow by substituting $f\to\bar f$ everywhere. More details on 
this method can be found in \citeres{di94,bo93}, where it is applied to 
$\Pem\gamma\to\Pem\gamma\gamma,\Pem\PZ\gamma$.

\paragraph{IR and collinear phase-space slicing}

Instead of using effective collinear factors, one can also apply
phase-space slicing to the collinear singularities, i.e.\ the collinear
regions are excluded by the angular cuts
\mbox{$\Delta\alpha<\alpha_f,\alpha_{\bar f}$} with $\Delta\alpha\ll 1$.
The integration over the collinear regions is particularly simple for
final-state radiation (see also \citeres{di94,bo93}), since collinear photon 
emission does not affect the kinematics in the factorized Born cross 
section $\rd\sigma_\born$ of the non-radiative process $\AAff$. 
The corrections from collinear photon emission can thus be described by
correction factors $\delta^\pm_\coll$ to $\rd\sigma_\born$,
\beq
\rd\sigma_\coll(\sigma,\bar\sigma) = 
2\delta^+_\coll\rd\sigma_\born(\sigma,\bar\sigma)+
\delta^-_\coll\rd\sigma_\born(-\sigma,\bar\sigma)+
\delta^-_\coll\rd\sigma_\born(\sigma,-\bar\sigma),
\eeq
where
\beqar
\delta^+_\coll &=& Q_f^2\frac{\alpha}{2\pi}
\left\{ 
\left[\ln\left(\frac{\Mf^2}{\Delta\alpha^2 E^2}\right)+1\right]
\left[2\ln\left(\frac{\Delta E}{E}\right)+\frac{3}{2}\right] 
+\frac{5}{2}-\frac{2\pi^2}{3} \right\},
\nn\\
\delta^-_\coll &=& Q_f^2\frac{\alpha}{4\pi}.
\eeqar
Note that the sum of the soft and collinear 
corrections without spin flip, i.e.\ 
$\delta_\soft+2\delta^+_\coll$, comprises all IR- and mass-singular terms
originating from real photon emission; after adding $\delta_\QED^\virt$,
all $\ln\lambda$ and $\ln\Mf$ terms drop out. On the other hand, the
corrections due to $\delta^-_\coll$ are the only sources for final-state
$f\bar f$ pairs with $\sigma=\bar\sigma$.

\paragraph{Subtraction method}

The idea of the subtraction method is to subtract and 
to add a simple auxiliary function from the singular integrand. 
This auxiliary function has to be chosen such that it cancels all 
singularities of the original integrand so that the phase-space 
integration of the difference can be performed numerically.  
Moreover, the auxiliary function has to be simple enough so that it can
be integrated over the singular regions analytically, when the
subtracted contribution is added again.
In the following we apply a modification of the so-called ``dipole
formalism'' \cite{ca96}, which was formulated for
next-to-leading-order QCD corrections involving 
unpolarized massless
partons. In the modified version of this formalism all divergences are
regularized by photon and fermion masses, and polarization is allowed
\cite{diprep}.

When the dipole formalism is applied to photon radiation, the
combinatorial part in the construction of the subtraction function is
rather simple. The subtraction function consists of contributions
labelled by all ordered pairs of charged external particles, one of
which is called {\it emitter}, the other one {\it spectator}.
Specifically, for $\AAffA$ we get two contributions: in the first case
$f$ plays the role of the emitter and $\bar f$ the one of the spectator,
and vice versa in the second case. 
The two functions that are subtracted
from $\sum_{\lambda'}|\M_\gamma|^2$ in the phase-space integral are
explicitly given by
\beqar
|\M_{\sub,1}^{\lambda_1,\lambda_2,\sigma,\bar\sigma}|^2 &=& 
\frac{Q_f^2 e^2}{(p\cdot k')}\left[\frac{2}{1-z_1(1-y_1)}-1-z_1\right]
|\M_\born^{\lambda_1,\lambda_2,\sigma,\bar\sigma}(s,t_1,u_1)|^2,
\nn\\
|\M_{\sub,2}^{\lambda_1,\lambda_2,\sigma,\bar\sigma}|^2 &=& 
\frac{Q_f^2 e^2}{(\bar p\cdot k')}\left[\frac{2}{1-z_2(1-y_2)}-1-z_2\right]
|\M_\born^{\lambda_1,\lambda_2,\sigma,\bar\sigma}(s,t_2,u_2)|^2,
\label{eq:msub}
\eeqar
where $|\M_\born^{\lambda_1\lambda_2\sigma\bar\sigma}|^2$ are the squared
Born helicity amplitudes \refeq{eq:mborn} for $\AAff$. The auxiliary variables 
$y_i$ and $z_i$ ($i=1,2$) are defined by
\beq
y_1 = \frac{pk'}{p\bar p + pk'+ \bar pk'} = \frac{2pk'}{s}, \qquad
z_1 = \frac{p\bar p}{p\bar p + \bar pk'}, \qquad
y_2 = y_1\Big|_{p\leftrightarrow\bar p}, \qquad
z_2 = z_1\Big|_{p\leftrightarrow\bar p}, 
\eeq
and the Mandelstam variables $t_i$ and $u_i$ are defined as in 
\refeq{eq:stu}, but for auxiliary momenta $p_i$ and $\bar p_i$,
\beqar
t_i &=& (k_1-p_i)^2 = (k_2-\bar p_i)^2 = -4E^2\sin^2\text\frac{\theta_i}{2}, 
\nn\\
u_i &=& (k_1-\bar p_i)^2 = (k_2-p_i)^2 = -4E^2\cos^2\text\frac{\theta_i}{2}.
\label{eq:tiui}
\eeqar
The auxiliary momenta are chosen such that $p_i\to p$ and 
$\bar p_i\to\bar p$ in the IR limit $k'\to 0$,
that $p_1\to p+k'$ and $\bar p_1\to \bar p$ if $k$ becomes collinear
to $p$, and that $\bar p_2\to \bar p+k'$ and $p_2\to p$ if $k$ becomes
collinear to $\bar p$. 
Moreover, the auxiliary momenta obey momentum conservation, $p+\bar
p+k'=p_i+\bar p_i$, and the mass-shell conditions,
$p_i^2=\bar p_i^2=0$,
\beqar
p_1 &=& \parbox{4.0cm}{$\disp p+k'-\frac{y_1}{1-y_1}\bar p,$} 
\bar p_1 = \frac{1}{1-y_1}\bar p,
\nn\\
p_2 &=& \parbox{4.0cm}{$\disp \frac{1}{1-y_2}p,$} 
\bar p_2 = \bar p+k'-\frac{y_2}{1-y_2}p. 
\eeqar
>From this definition we can also deduce that the scattering angles
$\theta_i$, which are defined in \refeq{eq:tiui}, are given by
$\theta_1=\theta_{\bar f}$ and $\theta_2=\theta_f$.

It is straightforward to check that $\sum_i|\M_{\sub,i}|^2$ 
has the same asymptotic structure as $\sum_{\lambda'}|\M_\gamma|^2$ in
the soft limit $k'\to 0$ and in the collinear limits 
$(p\cdot k'),(\bar p\cdot k')\to 0$,
so that the phase-space integral
\beqar\label{eq:int1}
\sigma_1 &=& \frac{\Ncf}{2s} \int \rd\Gamma \,
\biggl[\biggl(\sum_{\lambda'}|\M_\gamma|^2\biggr)
g_\cut(\theta_f)g_\cut(\theta_{\bar f}) 
-\biggl(\sum_i|\M_{\sub,i}|^2g_\cut(\theta_i) \biggr)\biggr]\\
 &=& \frac{\Ncf}{2s} \int \rd\Gamma \,
\biggl[\biggl(\sum_{\lambda'}|\M_\gamma|^2\biggr)
g_\cut(\theta_f)g_\cut(\theta_{\bar f}) 
-\biggl(|\M_{\sub,1}|^2g_\cut(\theta_{\bar f}) \biggr)
-\biggl(|\M_{\sub,2}|^2g_\cut(\theta_f) \biggr)\biggr] \nn
\eeqar
is finite and can be performed numerically. In \refeq{eq:int1} we
indicated the phase-space cuts \refeq{eq:aaffacut} explicitly by the 
step functions \refeq{eq:step}. In the subtracted part the cuts are
applied to the auxiliary momenta $p_i$, $\bar p_i$. Since these
approach the physical momenta of the 
final-state fermions in the singular regions, the cuts 
do not obstruct the cancellation of the
singularities in \refeq{eq:int1} as long as they avoid the
singularities. The last equality holds because $\theta_2=\theta_f$ and
$\theta_1=\theta_{\bar f}$ in our case.
 
For the full cross section we have to 
add the integral of $\sum_i|\M_{\sub,i}|^2$ that is evaluated with the 
regulators $\lambda$ and $\Mf$ \cite{diprep}. 
The functions $|\M_{\sub,i}|^2$ are constructed such that the integration 
over the photon phase space can be performed analytically, leading to
universal correction factors $\delta^\pm_\sub$ on the Born cross section
$\sigma_\born$ \refeq{eq:intborn} for $\AAff$,
\beq
\sigma_2(\sigma,\bar\sigma) = 
2\delta^+_\sub\sigma_\born(\sigma,\bar\sigma)+
\delta^-_\sub\sigma_\born(-\sigma,\bar\sigma)+
\delta^-_\sub\sigma_\born(\sigma,-\bar\sigma),
\label{eq:sigma2}
\eeq
where
\beqar
\delta^+_\sub &=& Q_f^2\frac{\alpha}{2\pi}
\left\{ 
\ln\left(\frac{\lambda^2}{s}\right)\ln\left(\frac{\Mf^2}{s}\right)
+\ln\left(\frac{\lambda^2}{s}\right)
-\frac{1}{2}\ln^2\left(\frac{\Mf^2}{s}\right)
+\frac{1}{2}\ln\left(\frac{\Mf^2}{s}\right)
+\frac{5}{2}-\frac{2\pi^2}{3} \right\},
\nn\\
\delta^-_\sub &=& Q_f^2\frac{\alpha}{4\pi}.
\eeqar
In \refeq{eq:sigma2} the Born cross sections $\si_\born$ are evaluated
with the restriction \refeq{eq:aaffcut} on the scattering angle $\theta$:
$\theta_\cut<\theta<180^\circ-\theta_\cut$. 
As required, the IR- and mass-singular terms in $2\delta^+_\sub$ exactly
cancel against those terms in $\delta_\QED^\virt$.
The final result for the real-photonic contribution to the cross section
is given by $\sigma_\gamma=\sigma_1+\sigma_2$.

\section{Numerical results}
\label{se:numres}

For the numerical evaluation we have adopted the parameters \cite{ca98}
\beq
\alpha = 1/137.0359895, \qquad
\MW = 80.41{\GeV}, \qquad
\MZ = 91.187\GeV.
\eeq
We need not specify the masses $\Mf$ of the light fermions, since these 
are only kept as regulating parameters and drop out in all considered
observables. 
We discuss only unpolarized cross sections. The non-vanishing cross
sections for polarized initial states and unpolarized final states
differ from the unpolarized cross sections only by the normalization.

\bfi
\centerline{
\setlength{\unitlength}{1cm}
\begin{picture}(12,8)
\put(0,0){\includegraphics{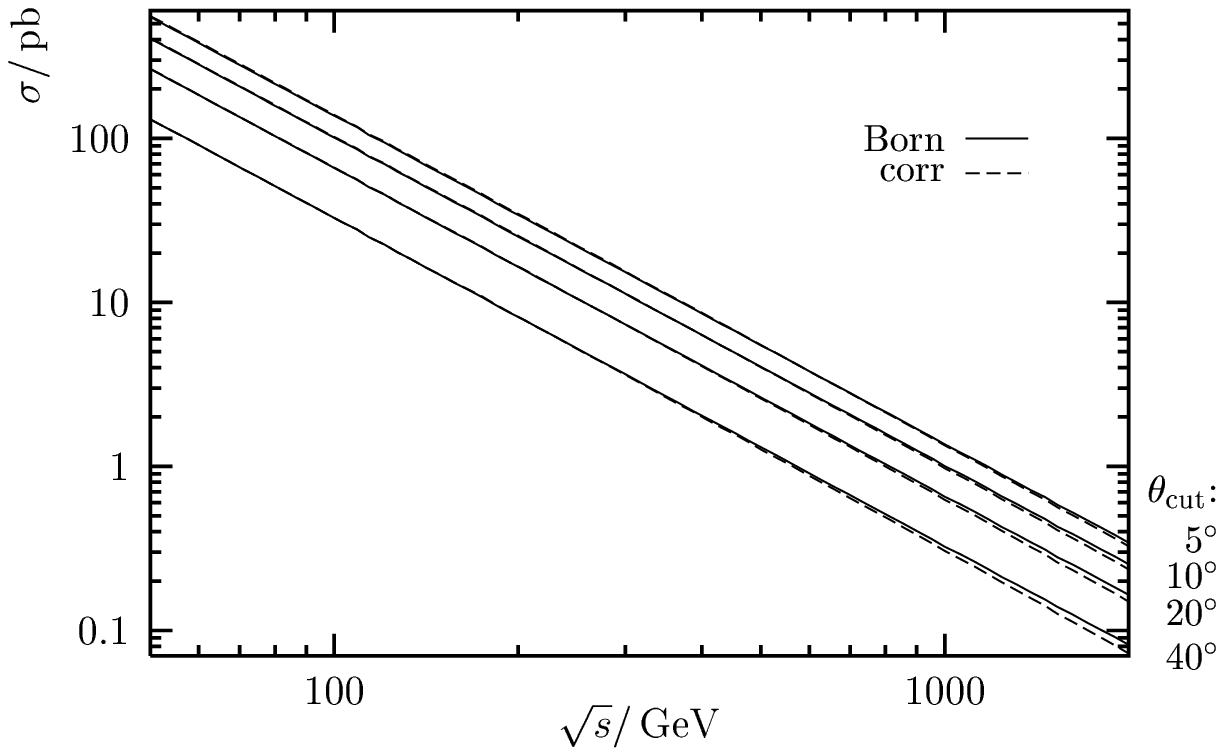}}
\end{picture}
} 
\caption{Lowest-order and ${\cal O}(\alpha)$-corrected cross section for
$\gamma\gamma\to\Pem\Pep$}
\label{fig:intcs}
\efi
In \reffi{fig:intcs} we show the lowest-order and the 
${\cal O}(\alpha)$-corrected cross section for $\AAee$ for the 
angular cuts $\theta_\cut=5^\circ,10^\circ,20^\circ,40^\circ$. 
The Born cross sections 
vary from $137\pba$ to $33\pba$ for these cuts at $\sqrt{s}=100\GeV$;
they scale like $1/s$ if the cut angle $\theta_\cut$ is
chosen energy-independent, as can be seen in \refeq{eq:intborn}.
Since the impact of the ${\cal O}(\alpha)$ corrections is hardly visible
in \reffi{fig:intcs},
we show the relative QED and weak corrections to
$\AAee$ separately in \reffi{fig:corr} for two angular cuts. 
\bfi
\centerline{
\setlength{\unitlength}{1cm}
\begin{picture}(12,16)
\put(0,8){\includegraphics{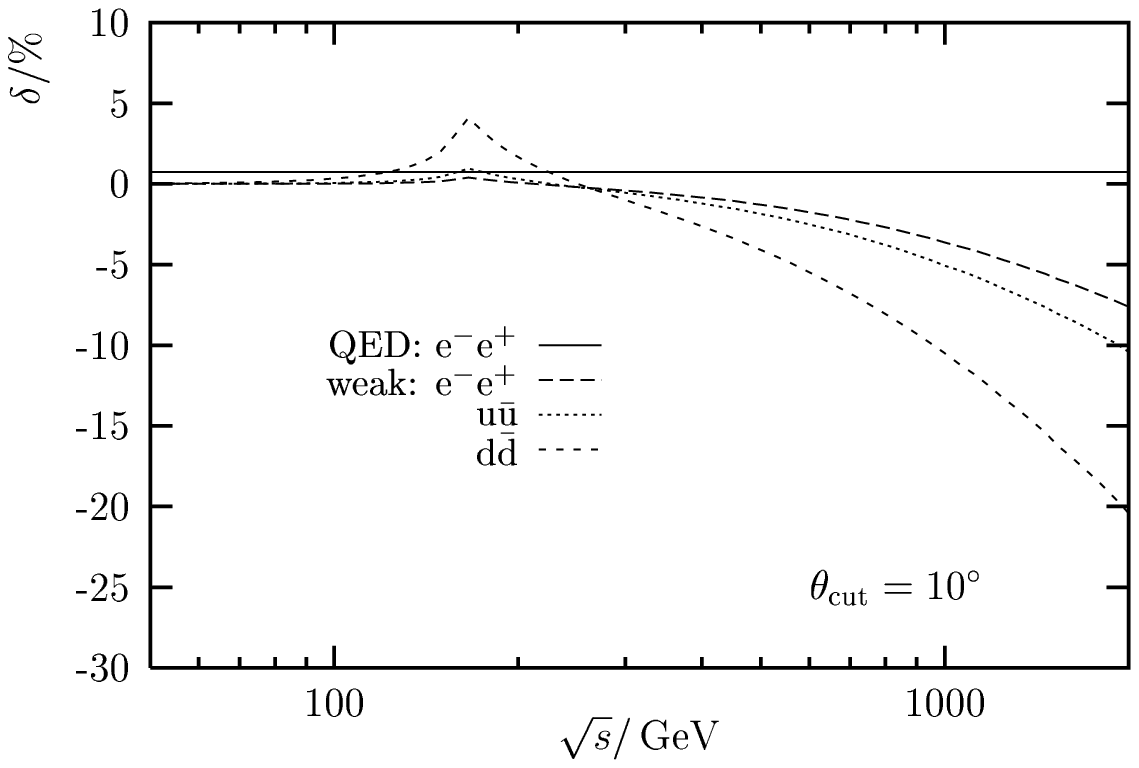}}
\put(0,0){\includegraphics{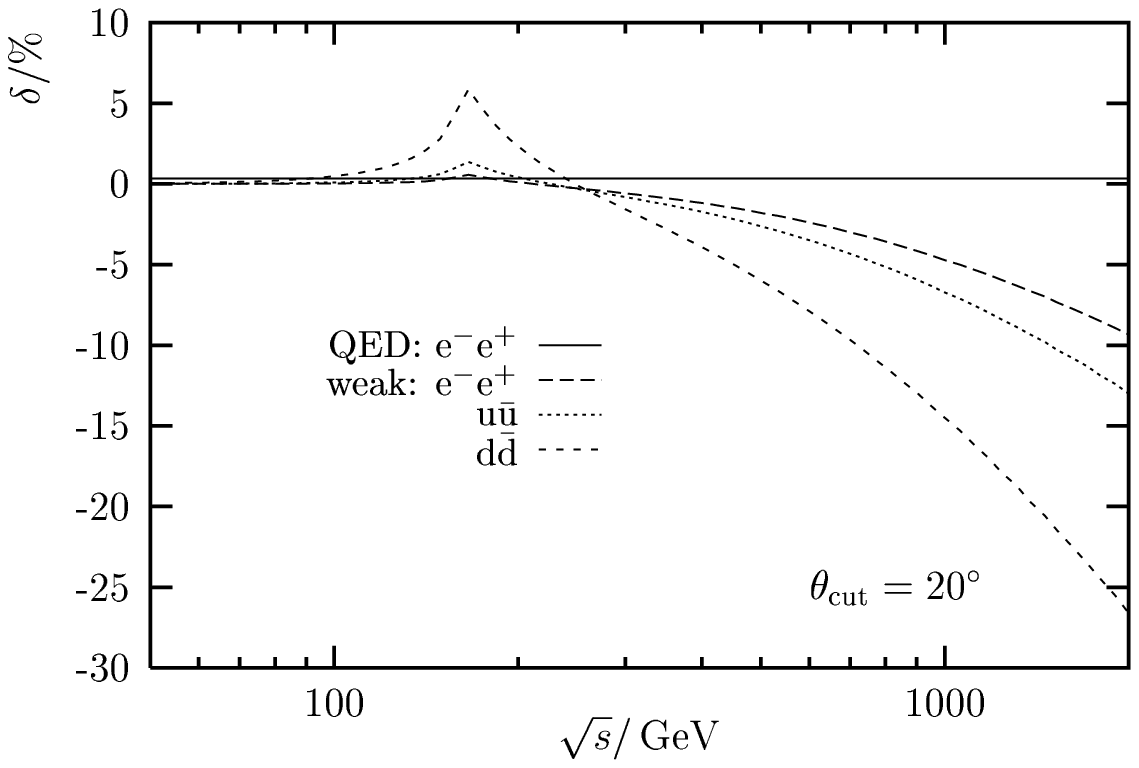}}
\end{picture}
} 
\caption{Relative QED and weak corrections to $\gamma\gamma\to\Pem\Pep$, 
and weak corrections to $\gamma\gamma\to\Pu\bar\Pu,\Pd\bar\Pd$}
\label{fig:corr}
\efi

For an energy-independent angular cut $\theta_\cut$, the QED corrections 
(see \reffi{fig:corr})
do not depend on the scattering energy for $s\gg\Me^2$, since all 
electron-mass singularities cancel, and $s$ is the only scale that survives. 
The cancellation of all potentially large QED corrections such as
$\alpha\ln(\Me^2/s)/\pi$ implies that the resulting QED correction is of
the order of ${\cal O}(\alpha/\pi)$, i.e.\ of the order of several per
mille. The numerical results confirm this expectation.
The weak corrections stay below $0.05\%$ for energies below
$100\GeV$ and tend to zero in the low-energy limit. In other words,
weak-boson exchange decouples below the electroweak scale. 
Above $100\GeV$, the weak corrections become sizeable and develop a peak
at $\sqrt{s}=2\MW$, originating from diagrams with a W-pair cut in the
$s$ channel. For energies up to $1\TeV$, 
$\delta_\weak$ reaches several per cent and becomes
more and more negative with increasing energy, crossing the $-10\%$ mark
at about $2\TeV$. The large negative corrections at high energies are
due to Sudakov-type logarithms such as $\alpha\ln^2(\MW^2/s)/\pi$, and the
dominant contributions stem from W-boson exchange.

The lowest-order cross sections and the relative QED corrections for
$\AAff$ with arbitrary fermion flavour can be easily obtained from the
results on $\AAee$ by multiplying the results for the $\Pem\Pep$ pair
by the factors $\Ncf Q_f^4$ and $Q_f^2$, respectively. In particular,
this means that the QED corrections to non-leptonic channels become
even smaller. The weak corrections, however, depend on the fermion
flavour in a non-trivial way. Therefore, the relative weak corrections
are explicitly shown in \reffi{fig:corr} also for up-type and
down-type light quarks. The shape of the weak corrections to
light-quark-pair production is qualitatively similar to the one for
lepton-pair production, but 
they are larger in size. For high energies we roughly
get $\delta_\weak^{\Pu\bar\Pu}/\delta_\QED^{\Pem\Pep} \sim 1.4$ and
$\delta_\weak^{\Pd\bar\Pd}/\delta_\QED^{\Pem\Pep} \sim 3$.
This enhancement of the relative weak corrections is mainly due to the
suppression of the lowest-order cross section by the quark charges,
which is not present in the dominating CC corrections.

\newcommand{\spc}{\phantom{$-$}}
\begin{table}
\begin{center}
\begin{tabular}{|c|c||c|c|c||c||c|}
\hline
$\sqrt{s}/\GeV$ & $\theta_{\mathrm{cut}}$ & 
$\sigma_\Born^{\Pem\Pep}/\pba$ & $\delta_\QED^{\Pem\Pep}/\%$ &
$\delta_\weak^{\Pem\Pep}/\%$ & $\delta_\weak^{\Pu\bar\Pu}/\%$ &
$\delta_\weak^{\Pd\bar\Pd}/\%$ 
\\ \hline\hline
  10 &  $5^\circ$ &  13722  & \spc 1.30 & \spc 0.00 & \spc 0.00 & \spc 0.00
\\ \cline{2-7}
     & $10^\circ$ &  10130  & \spc 0.74 & \spc 0.00 & \spc 0.00 & \spc 0.00
\\ \cline{2-7}
     & $20^\circ$ & 6595.2  & \spc 0.33 & \spc 0.00 & \spc 0.00 & \spc 0.00
\\ \cline{2-7}
     & $40^\circ$ & 3270.9  &   $-0.02$ & \spc 0.00 & \spc 0.00 & \spc 0.00
\\ \hline\hline
 100 &  $5^\circ$ & 137.22  & \spc 1.30 & \spc 0.02 & \spc 0.05 & \spc 0.26
\\ \cline{2-7}
     & $10^\circ$ & 101.30  & \spc 0.74 & \spc 0.02 & \spc 0.07 & \spc 0.34
\\ \cline{2-7}
     & $20^\circ$ & 65.952  & \spc 0.33 & \spc 0.03 & \spc 0.10 & \spc 0.49
\\ \cline{2-7}
     & $40^\circ$ & 32.709  &   $-0.02$ & \spc 0.05 & \spc 0.14 & \spc 0.73
\\ \hline\hline
 500 &  $5^\circ$ & 5.4889  & \spc 1.30 & $-0.97$   & $-1.39$   & $-3.03$
\\ \cline{2-7}
     & $10^\circ$ & 4.0520  & \spc 0.74 & $-1.29$   & $-1.85$   & $-4.08$
\\ \cline{2-7}
     & $20^\circ$ & 2.6381  & \spc 0.33 & $-1.78$   & $-2.62$   & $-5.97$
\\ \cline{2-7}
     & $40^\circ$ & 1.3084  &   $-0.02$ & $-2.47$   & $-3.79$   & $-9.23$
\\ \hline\hline
1000 &  $5^\circ$ & 1.3722  & \spc 1.30 & $-2.81$   & $-3.88$   &  $-7.95$
\\ \cline{2-7}
     & $10^\circ$ & 1.0130  & \spc 0.74 & $-3.61$   & $-5.03$   & $-10.49$
\\ \cline{2-7}
     & $20^\circ$ & 0.65952 & \spc 0.33 & $-4.70$   & $-6.69$   & $-14.48$
\\ \cline{2-7}
     & $40^\circ$ & 0.32709 &   $-0.02$ & $-5.95$   & $-8.78$   & $-20.11$
\\ \hline\hline
2000 &  $5^\circ$ & 0.34306 & \spc 1.30 & $-6.18$   &  $-8.27$  & $-15.92$
\\ \cline{2-7}
     & $10^\circ$ & 0.25325 & \spc 0.74 & $-7.62$   & $-10.35$  & $-20.38$
\\ \cline{2-7}
     & $20^\circ$ & 0.16488 & \spc 0.33 & $-9.33$   & $-12.98$  & $-26.60$
\\ \cline{2-7}
     & $40^\circ$ & 0.081773&   $-0.02$ & $-11.15$  & $-15.98$  & $-34.55$
\\ \hline
\end{tabular}
\end{center}
\caption{Integrated Born cross section for $\gamma\gamma\to\Pem\Pep$,
the corresponding relative QED and weak corrections, and the weak
corrections to $\gamma\gamma\to\Pu\bar\Pu,\Pd\bar\Pd$}
\label{tab:corr}
\end{table}
Table~\ref{tab:corr} summarizes the discussed results by providing some
representative numbers. 
At high energies the weak corrections are dominated by the CC
corrections. The NC corrections are at the level of 1\% at $2\TeV$.

We conclude our numerical discussion by a short comparison of the
different methods for the singular phase-space integration for real
photon emission, which are described in \refse{se:QEDreal}.
\begin{table}
\begin{center}
\begin{tabular}{|c||c|c||l@{$\,\pm\,$}l|l@{$\,\pm\,$}l|}
\hline
Method & $\Delta E/E$ & $\Delta\alpha/{\mathrm{rad}}$ & 
\multicolumn{2}{c|}{$\theta_{\mathrm{cut}}=10^\circ$} & 
\multicolumn{2}{c|}{$\theta_{\mathrm{cut}}=20^\circ$}
\\ \hline\hline
IR slicing and & $10^{-3}$ & -- & 0.798 & 0.016 & 0.345 & 0.014
\\ \cline{2-2}\cline{4-7}
effective collinear factor & $10^{-5}$ & & 0.819 & 0.029 & 0.329 & 0.024
\\ \hline\hline
IR and collinear slicing 
& $10^{-3}$ & $10^{-3}$ & 0.756 & 0.011 & 0.3302 & 0.0083
\\ \cline{3-7}
& & $10^{-5}$ & 0.784 & 0.015 & 0.349 & 0.013
\\ \cline{2-7}
& $10^{-5}$ & $10^{-3}$ & 0.734 & 0.019 & 0.323 & 0.015
\\ \cline{3-7}
& & $10^{-5}$ & 0.808 & 0.027 & 0.324 & 0.022
\\ \hline\hline
Subtraction scheme & -- & -- & 0.74447 & 0.00080 & 0.33124 & 0.00069
\\ \hline
\end{tabular}
\end{center}
\caption{Comparison of results for the QED correction $\delta_{\QED}/\%$ 
at $\protect\sqrt{s}=500\protect\GeV$,
obtained by the different methods for bremsstrahlung corrections
described in \protect\refse{se:QED} }
\label{tab:qed}
\end{table}
Table~\ref{tab:qed} compares numerical results on $\delta_\QED$ that have 
been obtained by performing the 
multidimensional integration with {\sl Vegas}
\cite{vegas}, using the same {\sl Vegas} parameters for each integration.
The subtraction method leads to an integration
error that is smaller by a factor of 10--20 with respect to the results
from the two versions of phase-space slicing. While there are still
large compensations between the phase-space integral and the
(semi-)analytically calculated singular parts in the slicing approach, for
the subtraction method all compensations take place between
$\delta_\QED^\virt$ and $2\delta^+_\sub$, which are computed without
delicate numerical integrations. Table~\ref{tab:qed} illustrates the
consistent application of the different methods, but the numbers for the
smaller cut $\theta_\cut=10^\circ$ also reveal
that Monte Carlo integration by {\sl Vegas} tends to underestimate integration
errors if the integrand becomes complicated, although it has
been smoothed by appropriate transformations of the integration
variables%
\footnote{Repeated evaluations for $\theta_\cut=10^\circ$ show that the 
results obtained with the two slicing variants come closer and closer
to that of the subtraction method given in \refta{tab:qed} if the
statistics is improved.}.
This distinguishes the subtraction method that
is less sensitive to numerical uncertainties.

\section{Summary}
\label{se:Sum}

The ${\cal O}(\alpha)$ corrections to $\AAff$ in the Standard Model 
have been calculated for arbitrary light fermions $f$, i.e.\
fermion-mass effects are neglected. Compact analytical results for 
the cross sections have been listed for arbitrary polarization 
configurations, rendering their incorporation in computer codes very
simple.
Numerical results on the corrections to integrated cross sections have
been discussed.

The corrections are classified into QED and purely weak corrections. 
Owing to the cancellation of all mass-singular contributions between
virtual and real-photonic corrections, the QED corrections to integrated
cross sections are of 
${\cal O}(Q_f^2\alpha/\pi)$ for all energies, i.e.\ of the order of 
some per mille. For lepton-pair production the weak corrections are 
negligible below the weak bosons scale, reach some per cent at $1\TeV$,
and reduce the cross section more and more with increasing energy, 
crossing $-10\%$ at about $2\TeV$. For up- and down-type quarks the weak
corrections to the integrated cross sections show the same qualitative
features as in the leptonic case, but the corrections are 
a few times larger. The weak corrections
vanish whenever the differential cross sections develop $t$- or
$u$-channel poles, i.e.\ the relative corrections can be enhanced or
suppressed by appropriate angular cuts.

The smallness and the structure of the corrections to 
$\AAee,\mu^-\mu^+$ underline the suitability of these processes as
a luminosity monitor. In particular, the corrections do not exhibit large
uncertainties due to hadronic effects in the photonic vacuum
polarization or due to the less precisely known top-quark or even
unknown Higgs-boson mass.
The results for the processes $\ga\ga\to q\bar q$ provide a 
valuable input for QCD studies.

\appendix
\def\theequation{\thesection.\arabic{equation}}
\setcounter{equation}{0}
\section*{Appendix}

\section*{List of scalar integrals}
\setcounter{section}{1}
\label{scalar}

Here we list all scalar one-loop integrals that
are needed for the evaluation of the virtual corrections 
given in \refse{se:EWRC}. We use the same definition of the
momentum-space integrals and of the arguments of the standard
functions $B_0$, $C_0$, and $D_0$ as given in the appendix of
\citere{di94}.  The relevant integrals are calculated for the limit
$|s|,|t|,|u|,\MW^2\gg\Mf^2,\Mff^2$.  By definition, Mandelstam
variables with a hat get an infinitesimal imaginary part $\ieps$, with
$\eps>0$, i.e.\ $\hat s=s+\ieps$ etc. After supplying this imaginary
part where necessary, all scalar integrals can also be obtained from
those for Compton scattering in \citere{di94} upon using crossing
symmetry.  Scalar functions that are related by the interchange of $t$
and $u$ are given generically with the abbreviation $r=t,u$.

All needed 2-point functions $B_0$ are calculated in $D$ 
space-time dimensions with $D\to 4$.
Instead of using $B_0$ directly, we have preferred to introduce the
UV-finite combinations
\beqar
B_{0}(s,0,0)-B_{0}(0,0,\MW) &=& B(s)         
= \ln\left(-\frac{\MW^2}{\hat s}\right)+1,
\nn \\[.5em]
B_{0}(r,0,\MW)-B_{0}(0,0,\MW) &=& B_{\Pw}(r)
= \left(\frac{\MW^2}{r}-1\right)\ln\left(1-\frac{\hat r}{\MW^2}\right)+1,
\nn \\[.5em]
B_{0}(s,\MW,\MW)-B_{0}(0,0,\MW) &=& B_{\Pw\Pw}(s)
= \beta_{\Pw}\ln(x_\Pw)+1,
\eeqar
with the abbreviations
\beq
x_{\Pw} = \frac{\beta_{\Pw}-1}{\beta_{\Pw}+1},
\qquad \beta_{\Pw} = \sqrt{1-\frac{4\MW^2}{\hat s}}.
\label{eq:xw}
\eeq

The relevant 3- and 4-point functions are given by
\beqar
C_{0}(0,0,s,\Mff,\Mff,\Mff) &=& C(s)=\frac{1}{2s}
\ln^2\left(-\frac{\hat s}{\Mff^2}\right),
\nn\\[.5em]
C_{0}(\Mf^2,0,r,\MW,\Mff,\Mff) &=& \bar C_{\Pw}(r)
= \frac{1}{r}\left[ \Li_2\left(\frac{r}{\MW^2}\right)
-\ln\left(\frac{\Mff^2}{\MW^2-r}\right)
\ln\left(1-\frac{r}{\MW^2}\right)
\right],
\nn\\[.5em]
C_{0}(0,0,s,0,\MW,0) &=& C_{\Pw}(s)= \frac{1}{s}\left[
-\Li_2\left(1+\frac{\hat s}{\MW^2}\right)+\frac{\pi^2}{6}
\right],
\nn\\[.5em]
C_{0}(0,0,r,0,\MW,\MW) &=& \bar C_{\Pw\Pw}(r)
= -\frac{1}{r}\Li_2\left(\frac{r}{\MW^2}\right),
\nn\\[.5em]
C_{0}(0,0,s,\MW,0,\MW) &=& C_{\Pw\Pw}(s) 
= \frac{1}{s}\ln^2(x_{\Pw}),
\nn\\[.5em]
C_{0}(0,0,s,\MW,\MW,\MW) &=& C_{\Pw\Pw\Pw}(s)
= \frac{1}{2s}\ln^2(x_{\Pw}),
\eeqar
\beqar
D_0(0,0,0,0,s,r,\Mff,\MW,\Mff,\Mff) &=& D_{\Pw}(s,r) 
= \frac{1}{s(r-\MW^2)}\left[
\Li_2\left(1+\frac{\hat s}{\MW^2}\right)
\right.
\nn\\[.3em]
&& \hspace{-15em} \left. {}
-4\Li_2\left(\frac{r}{r-\MW^2}\right)
+\frac{1}{2}\ln^2\left(-\frac{\hat s}{\Mff^2}\right)
+2\ln\left(-\frac{\hat s}{\Mff^2}\right)\ln\left(1-\frac{r}{\MW^2}\right) 
-\frac{\pi^2}{6}\right],
\nn\\[.5em]
D_0(0,0,0,0,t,u,\MW,\MW,\Mff,\Mff) &=& D_{\Pw\Pw}(t,u) 
=  \frac{1}{tu-\MW^2(u+t)} 
\nn\\[.3em]
&& \hspace{-15em} \times\left[
2\Li_2\left(1+x_{tu}-\frac{\hat t}{\MW^2}x_{tu}\right)
+2\eta\left(-x_{tu},1-\frac{\hat t}{\MW^2}\right)
\ln\left(1+x_{tu}-\frac{\hat t}{\MW^2}x_{tu}\right)
\right.
\nn\\[.3em]
&& \hspace{-15em} \phantom{\times\biggr\{}\left. {}
-2\Li_2(1+x_{tu})
+\ln\left(\frac{\MW^2-t}{\Mff^2}\right)
\ln\left(1-\frac{t}{\MW^2}\right)
\right] \;+\; (t\leftrightarrow u),
\nn\\[.3em]
&& \hspace{-15em} \mbox{with } \quad
x_{tu} = \frac{\MW^2(\hat t+\hat u)}{\hat t\hat u-\MW^2(\hat t+\hat u)},
\nn\\[.5em]
D_0(0,0,0,0,s,r,\MW,0,\MW,\MW) &=& D_{\Pw\Pw\Pw}(s,r) 
\nn\\[.3em]
&& \hspace{-16em} 
= \frac{1}{\sqrt{\hat s^2(\hat r-\MW^2)^2-4\hat r^2\hat s\MW^2}} 
\;\disp\sum_{n=1}^{2}(-1)^{n+1}
\nn\\[.3em]
&& \hspace{-15em} \times\left[
3\Li_2(1+x_n)
-\Li_2\left(1+\frac{x_n\MW^2}{\MW^2-\hat r}\right)
-\eta\left(-x_n,\frac{\MW^2}{\MW^2-\hat r}\right)
\ln\left(1+\frac{x_n\MW^2}{\MW^2-\hat r}\right)
\right.\nn\\[.3em]
&& \hspace{-15em} \phantom{\times\biggr\{}\left. {}
+\ln\left(1-\frac{\hat r}{\MW^2}\right)\ln(-x_n)
-\sum_{\tau=\pm 1} \left\{
\Li_2\left(1+x_n x_{\Pw}^\tau\right)
+\eta\left(-x_n,x_{\Pw}^\tau\right) \ln\left(1+x_n x_{\Pw}^\tau\right) 
\right\} \right],
\nn\\
\eeqar
with $x_\Pw$ and $\beta_\Pw$ as given in \refeq{eq:xw} and
\beq
x_{1,2} = \left[\hat s(\hat r-\MW^2)-2\hat r \MW^2
\pm\sqrt{\hat s^2(\hat r-\MW^2)^2-4\hat r^2\hat s\MW^2}\right]
/[2(\hat r+\hat s)\MW^2].
\eeq
The dilogarithm $\Li_2(x)$ and the function $\eta(x,y)$ are defined
as usual:
\beqar
\Li_2(x) = -\int_0^x\,\frac{\rd t}{t}\,\ln(1-t), 
&&\qquad -\pi<{\mathrm{arc}}(1-x)<\pi,\\[.3em]
\eta(x,y) = \ln(xy)-\ln(x)-\ln(y),
&&\qquad -\pi<{\mathrm{arc}}(x),{\mathrm{arc}}(y)<\pi.
\eeqar

\end{document}